# A New 3D Geometric Approach to Focus+Context Lens Effect Simulation

Bo Li, Xin Zhao and Hong Qin *
Stony Brook University

## Abstract

We present a novel methodology based on geometric approach to simulate magnification lens effects that can be utilized for Focus+Context (F+C) visualization. Our aim is to promote new applications of powerful geometric modeling techniques in visual computing. Conventional image processing/visualization methods are computed in two dimensional space (2D). we examine this conventional 2D manipulation from a completely innovative perspective of 3D geometric processing. Compared with conventional optical lens design (such as fish-eyes, bi-focal lens), 3D geometric method are much more capable of preserving shape features and minimizing distortion. This paper's novelty lies at integrating geometric deformation into lens design and taking a brand new geometric approach to solve the visualization problem. After identifying regions of interest (ROIs), we first embed the 2D screen into 3D space, and then equip it with a special 3D triangle mesh. We magnify an area of interest to better visualize the interior details, while keeping the rest of area without perceivable distortion. Second, we flatten the mesh back into 2D space for viewing, and further applications in the screen space. In both steps, we devise an iterative deformation scheme to minimize distortion around both focus and context regions, while avoiding the noncontinuous transition region between the focus and context areas. Particularly, our method allows the user to flexibly modify the ROI shapes to accommodate complex feature. The user can also easily specify a spectrum of metrics for different visual effects. Various experimental results demonstrate the effectiveness, robustness, and efficiency of our framework.

CR Categories: I.3.3 [Computer Graphics]: Three-Dimensional Graphics and Realism—Display Algorithms

## 1 Introduction

In recent years, we have witnessed the rapid advances in data acquisition, GPU rendering, and internet bandwidth. There is a stronger-than-ever need for visualizing large-scale datasets in various science/engineering applications. Meanwhile, with the explosive emergence of various types of portable devices (e.g., smartphone), the industry frequently pursues as-large-as-possible data visualization on physically-small-sized screen of mobile device in recent years. Therefore, a careful tradeoff is required to deal with the potentially conflicting requirement of the inherent screen size limitation and ever-increasing data size. Focus+Context visualization offers a good strategy when tackling this problem.

*e-mail:{bli, xinzhao, qin}@cs.stonybrook.edu

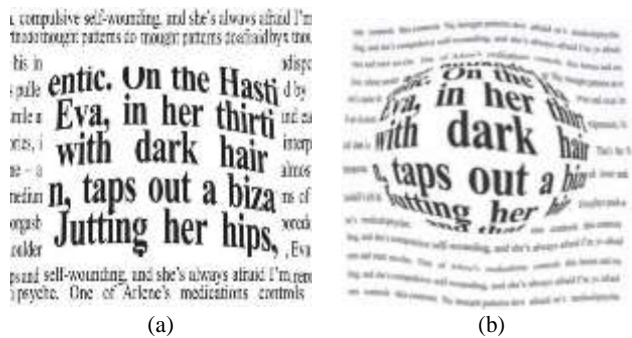

Figure 1: (a): Direct zoom-in. (b): Our geometric approach to simulate magnification lens

The traditional method is direct zoom-in, as shown in Fig. 1(a). However, the useful information is occluded after magnifying the focus region. Focus+Context (F+C) visualization (Fig. 1(b)), as a natural solution, has gained much research momentum recently. In order to display regions of interest (ROIs) with high resolution, F+C allows the user to access and address the detail of interest ("Focus") while still keeping the overall content of the whole data to accommodate human cognitive custom ("Context"). For example, when we observe the map for road planning, we need to focus on the local road name ("Focus"). However, the user may still prefer to know the connection from the local road to other regions ("Context"). Directly zoom-in the local road will occlude the road connection.

An attractive F+C visualization should consider many quality-related aspects: Shape preserving plays a crucial role during magnification. To improving the visual cognition, we need to preserve the shape of both focus and context regions. The improper magnification distortion may cause serious cognitive confusion; Meanwhile, smooth transition between focus region and context region is necessary. Any visual gain from unifying the detail with the surrounding context may easily be lost if the transition between the focus and context regions is difficult to understand; Also, an ideal system should provide flexibility to the user. For example, to handle data with complex and multiple ROIs, the user may have preference to use different magnification lens or different shapes of lens with changeable visual effects.

It is a tremendous challenge to optimize the output simultaneously with respect to all of these criteria. For example, many recent methods attempted to simulate optical lenses in depth (e.g., fisheyes, bifocal lens) for magnification. The most challenging side effect is that, it rarely considers shape-preserving and smooth transition, thus lens distortions are intolerable when features become sufficiently intricate.

Our new idea is to address the lens design and simulation problem using geometric approach. To provide more flexibility and enhance the ability to control distortion, we examine this conventional 2D deformation task from a completely innovative perspective of 3D geometric processing. Rather than manipulating on a 2D image/grid, we transform the 2D input to 3D mesh, and then conduct 3D deformations to magnify ROIs. To achieve our goal, we design a novel deformation framework that functionally acts as a "lens". We first build a special 3D triangle mesh that magnifies any area of interest while keeping the rest of area unchanged. Then, we automatically deform the 3D mesh back into 2D space for viewing. Both steps require us to find distortion minimization for each individual mesh element.

This paper presents a general theoretical and computational framework, in which 3D geometric modeling techniques can be systematically applied to the 2D lens simulation. The main contributions of our lens design and simulation include: (1) Our algorithm minimizes the geometric deformation metric distortion thus it is particulary suitable to satisfy the shape preserving property. Moreover, our deformation scheme lets the deformable mesh locally confine the resulting distortion with great flexibility rather than letting the distortion uniformly spread throughout the nearby spaces; The resulting transition between the focus and context regions is also smooth and seamless; (2) Instead of only using lenses with a regular circle or square shape, our system allows to design an arbitrary shape of magnifiers using to adapt complex shapes; (3) The user can iteratively specify the geometric metrics, which allows easy production of different visually pleasing effects. The whole algorithm is shown to be of high efficiency, because of the computation of a linear system with pre-processing acceleration.

Our paper is structured as follows. In Section 2, we overview related work. In Section 3, we briefly introduce our framework. We detail our algorithm of mesh magnification in Section 4 and discuss flattening in Section 5. We describe the implementation details and comparisons in Section 6. Section 7 concludes the paper with discussion on future work.

## 2 Previous Work

Various F+C visualization techniques have already been proposed on many applications. Plaisant et al. [Plaisant et al. 2002] defined the SpaceTree as a novel tree browser to support exploration in the large node link tree. The algorithm applies dynamic re-scaling of branches to best fit the space and includes integrated search and filter functions. For the seamless F+C, Shi et al. [Shi et al. 2005] proposed a distortion algorithm that increases the size of a node of interest while shrinking its neighbors. Gansner et al. [Gansner et al. 2004] presented a topological fisheye view for the visualization of large graphs. A method to cope with map and route visualization is proposed by Ziegler et al. [Ziegler and Keim 2008]. They depicted navigation and orientation routes as a path between nodes and edges of a topographic network. Recently, Karnick et al. [Karnick et al. 2010] presented a novel multifocus technique to generate a printable version of a route map that shows the overview and detail views of the route within a single, consistent visual frame. Different from the above methods with specific pre-defined targets, our framework is capable of handling various information or visualization-based applications.

The key component in F+C visualization is to design an efficient lens. Optical effects, such as fisheye [Furnas. 1986] for the non-linear magnification transformation with multi-scale, have been widely used. Fisheye views can enlarge the ROI while showing the remaining portions with successively less detail. Fisheye lens offers an effective navigation and browsing device for various applications [Nekrasovski et al. 2006]. In addition, InterRing proposed by Yang et al. [Yang et al. 2003] and Sunburst proposed by Stasko et al. [Stasko and Zhang 2000] have incorporated multi-focus fisheye techniques as an important feature for radial space-filling hierarchy visualization. The major advantage of the fisheye lens is the ability to display the data in a continuous manner, with a smooth transition between the focus and context regions. Although fisheye lens has advantages in preserving the spatial relation, it creates noticeable distortions towards its edges, which fails to formally control the focused region and preserve the shape features in the context region.

Aiming to cope with the shortcomings of the basic fisheye lens, more sophisticated lenses have been proposed. Bier et al. [Bier et al. 1993] presented a user interface that enhances the focal interest features and compresses the less interesting regions using a Toolglass and Magic Lenses. Carpendale et al. [Carpendale et al. 1997] proposed several view-dependent distortion patterns to visualize the internal ROI, where more space is assigned for the focal region to highlight the important features. LaMar et al. [LaMar et al. 2001] presented a fast and intuitive magnification lens with a tessellated border region by estimating linear compression according to the radius of lenses and texture information. Pietriga et al. [Pietriga and Appert 2008] provided a novel sigma lens with new dimensions of time and translucence to obtain diverse transitions. Later, they provided in-place magnification without requiring the user to zoom into the representation and consequently lose context [Pietriga et al. 2010]. Their representation-independent system can be implemented with minimal effort in different graphics frameworks. Meanwhile, the deformation methods are recently used for the complicated 3D datasets, including volume data [Correa and Silver 2007] [Wang et al. 2011] and mesh model [Wang et al. 2008]. Wang et al. [Wang et al. 2008] presented a method for magnifying features of interest while deforming the context without perceivable distortion, using an energy optimization model for large surface models. Later, they further extended this framework into 3D volumetric datasets [Wang et al. 2011, Wang et al. 2012]. However, the global shape distortion, mentioned as one of their major limitations, is issue for their system. By comparison, our technique renders the result keep- ing upper/lower body proportion, without obvious shape confusion that may have negative influences on the accuracy of object cog- nition. In addition, we utilize geometric deformation that applies to visualization of 2D data sets, targeting to eliminate the local an- gle distortion and keep the visual continuity. By comparison with Wang et al. [Wang et al. 2008], our framework supports more flex- ible metric design (see Section 5.1) to satisfy various requirements. Meanwhile, because we focus on processing the informatics data, multi-scale details are revealed after the magnification rather than the simple interpolation.

Many image deformation techniques have been successfully studied and used for various image manipulation applications like image editing and resizing. For example, Schaefer et al. [Schaefer et al. 2006] utilized moving least squares to fit transformations and achieve image editing. Also, many blending polynomial coordinates have been developed for better shape interpolation with boundary deformation constraints (e.g., biharmonic weights [Jacobson et al. 2011], green coordinates [Lipman et al. 2008]). Also, we observe the fact that all of the above techniques confine their operations as energy minimization in the 2D space only. Therefore, it is very attractive in this paper to explore a new deformation method that utilizes 3D geometric modeling techniques and broaden the scope of geometric modeling to help the visualization process.

## 3 Framework

This section gives a high level overview of our proposed framework. Our system takes as input a ready-to-display 2D image. For 3D dataset (e.g., volume datasets and 3D scanning models), we can generate the 2D format image through volume rendering. In geometric deformation, we can consider our input as a 2D regular triangle mesh $M = \{V, E, T\}$. $T = t_1, t_2, \ldots, t_n$ denotes every individual triangle, and $\{E, V\}$ denotes the sets of edges and vertices. Each vertex $v_i = (p_i, \varphi_i)$ includes the vertex 2D position $p_i = (x_i, y_i)$ and texture mapping coordinate $\varphi_i = (x_i, y_i)$. Note that in the input mesh the vertex position and mapping coordinates have the same value. The output is also a 2D triangle $M^{out}$ which has the same structure as $M$, but every vertex's position and mapping coordinate are updated. Fig. 2 illustrates our framework step-by-step using a google map as the example. Our framework mainly includes the following steps.

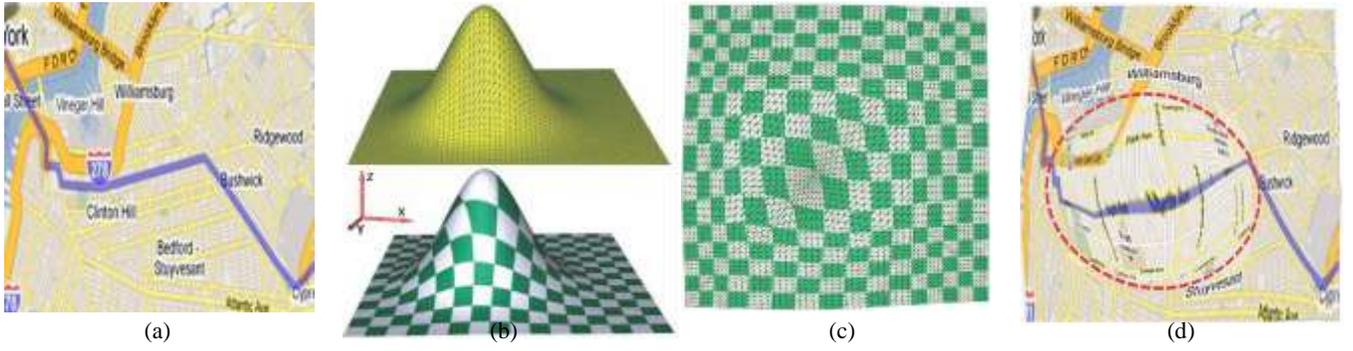

Figure 2: An example of our entire framework: (a) The input image. (b) We generate a 3D mesh to magnify the area of ROI. Then we transfer the texture from the input to the 3D mesh. (c) We deform the 3D mesh back into a 2D plane with minimized distortion. (d) Finally we get a new 2D image with area of ROI magnified.

Step 1. The user makes an initial choice about regions of interest (ROIs). We can use a simple user sketch (e.g., drawing a circle) as the ROI boundary to enclose each ROI, or use the exact shape/boundary of every ROI. The boundary can be determined by an automatic feature segmentation operation such as [Shi and Malik 1997] or simple heuristic methods.

Step 2. Generate a 3D mesh $M^{3D}$ based on the initial mesh $M$ in order to magnify the area of mesh on ROI.

- (2.1) For each ROI, we deform the original 2D surface patch in ROI into a specified 3D surface, with the ROI boundary as constraints (no shape changes outside the boundary). Every triangle's area in the boundary is therefore magnified.
- (2.2) We transfer the texture from $M$ to $M^{3D}$ while satisfy- ing the shape preserving property. To achieve this, for each
 vertex inside ROI boundary we compute texturing mapping coordinates $[u, v]$ on $M^{3D}$ by solving the harmonic equation $\nabla_2 u = 0$ and $\nabla_2 v = 0$.

Step 3. We deform $M^{3D}$ back into a 2D plane with distortion minimization. We flatten each triangle $t_i^{3D}$ in $M^{3D}$ back to 2D by rotation, and we denote this 2D triangle as standard triangle $t_i^{std}$. To make each triangle in the final output $M^{out}$ approximate to its standard triangle, we design an iterative-executed algorithm with two phases: For each iteration k, we have a starting 2D triangle mesh $M^k$ which is the result from (k − 1)th iteration ($M^0$ is initialized by projecting $M^{3D}$ to 2D).

- (3.1) For each triangle $t_i$ in $M^k$, we compute a deformation metric $M_i$ (formulated as a $2 \times 2$ matrix) using the standard triangle $t_i^{std}$.
- (3.2) We determine the updated position of every vertex by solving the linear equation to approximate the deformation metric $M_i$ for each triangle.

## 4 Mesh Magnification

The input of our framework is the uniform 2D dataset. Aiming to effectively generate the 2D rendered image from the mesh model/volumetric dataset, we adapt the fragment program (initially proposed by Stemaier et al. [Stegmaier et al. 2006]) for rendering, considering many parameters including depth, view angle, and camera position. The steps include: cast the ray into the mesh model/volume dataset and composite the color based on the surface/volume data and transfer functions, and render the result into the frame buffer for display.

In most practical focus+context visualization applications, the user only chooses a general approximate region via simple user sketch and/or basic geometric primitives (like the region within a drawn circle), enclosing both mesh segment and nearby context space as a reasonable proxy. The choice of circle lens is natural and humans are more accustomed to it with better visual understanding compared with other geometric primitives. In practice, we first visually choose a general/approximate region, then we pick the center c of this region as the center of sphere associated with a radius r. r must be large enough to enclose the entire ROI.

After we setting the lens, we magnify it by moving each vertex to a 3D position. we use gaussian function to compute $z_i$ for each vertex: $z_i = g(1 − \frac{d_i}{r})h_0$, where $d_i$ denotes the distance to the circle center c, g(x) denotes a standard gaussian function $e^{x^2}$ and $h_0$ is a user input to scale the magnification; As an alternative solution, we can also use a standard sphere instead of gaussian function to accommodate user's visual preference: $z_i = \sqrt{r^2 − d^2}$.

Arbitrary ROI boundary design. Our system also allows an exact boundary of an object in the image as the ROI boundary. We denote the triangle mesh patch inside this object as $M_p$ and $\partial M_p$ as the patch boundary. We first conduct the medial axis transform for $M_p$, generating a central curved path $C$ and each vertex $v_i$ in $M_p$ has a distance $d_i$ as the shortest distance to the path. The user decides the height $h_0$ of curved path $C$. For each vertex $v_i$, we have its new position $(x_i, y_i, z_i)$, $z_i = g(1 − \frac{d_i}{d_m})h_0$, where $d_m$ is the maximum distance. We need to subdivide the triangle if it is scaled or sheared too much after magnification. Then we interpolate the locations, colors, distances and heights linearly for newly-inserted vertices.

The automatic algorithm can handle versatile models very well, sometimes users still prefer to use special shapes as the desirable lenses for ROI. Fig. 3 shows different visual effects with different meshes.

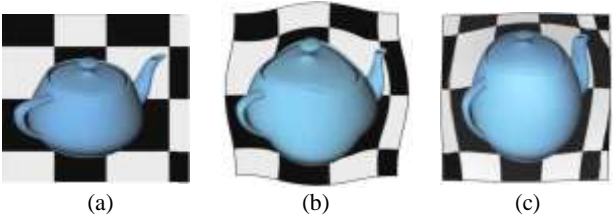

Figure 3: Magnification results using different shapes of lenses for the 3D teapot mesh model. (a) Original teapot mesh model. (b-c) Magnification results using the square-shaped and our automatically-generated ROI-guided meshes, respectively.

### 4.1 Texture Mapping

The objective of this step is to assign the texture to the magnified 3D triangle mesh, otherwise the texture will be distorted after changing every triangle's shape inside ROIs.

Since both the input mesh $M$ and magnified mesh $M^{3D}$ have squared boundary, we treat this problem as the energy minimization problem. We shall map the mesh $M^{3D}$ to a uniform 2D domain by solving the harmonic functions $\nabla^2 u = 0$ and $\nabla^2 v = 0$, where $\nabla^2 = \frac{\partial^2}{\partial x^2} + \frac{\partial^2}{\partial y^2}$. In practice, solving equations for any but the simplest geometries must resort to an efficient approximation due to the lack of closed-form analytical solutions in the general setting, we shall use mean value coordinates [Floater 2003] to solve it numerically.

- We assign each vertex an initial coordinate. In practice we initialize it with its original 2D position $(u_i, v_i) = (x_i, y_i)$.

- We iteratively update the coordinates for each vertex $(u_i, v_i) = \sum_{Ng(v_i)} w_j (u_j, v_j)$, and $N g(v_i)$ is the one-ring neighbor of $v_i$, $(u_j, v_j)$ is a neighbor's coordinate, $w_j$ is the local mean value coordinate [Floater 2003] computed on $M^{3D}$. Two types of vertices serve as the Dirichlet boundary conditions (i.e., we avoid changing their coordinates): (1) The squared boundary only; (2) All regions outside any ROI.

## 5 Flattening

We search for a flattened mesh so that we can display the result on the popular flat screen (Note that, our algorithm also supports curved screen like "IMAX"). The key challenge in this problem is to preserve the important geometric deformation metric for each triangle. The shape distortion can be measured as the total differences between the resulting triangles and the original triangles. We use the following algorithm to minimize the differences.

Step 1. For each triangle $t_i$ in 3D space, we reformulate it into a standard 2D triangle $t_i^{std}$ which keeps its original shape. Suppose $v_1$, $v_2$, $v_3$ are 3 vertices of $t_i^{3D}$ in 3D space, $e_1 = v_1 - v_2$, $e_2 = v_2 - v_3$, $e_3 = v_1 - v_3$ are 3 edge vectors. We recompute 2D positions of 3 vertices as $v_1 = (0, 0)$, $v_2 = (\|e_1\|, 0)$ and $v_3 = (\|e_2\|\cos\theta, \|e_2\|\sin\theta)$ (Fig. 4). $\theta$ is the angle between $e_1$ and $e_2$.

Note that, we flatten the triangle separately so a vertex in $M$ has different 2D positions in different $t_i^{std}$.

Step 2. Now we flatten the mesh back to 2D. This step includes 2 iteratively computed phases. The output mesh $M^{out}$ has the same triangle mesh structure as $M$ while every vertex has only a 2D position. Initially, we guess $M^0 = M^{3D}$ and we reduce the dimension of vertices to 2D by projecting along axis-z: $v_i = (x_i, y_i)$.

(2.1) In this phase we compute the deformation metric for each triangle. The metric represents the transformation from the localized standard $t_i^{std}$ to its k-th iteration counterpart $t_i^k$. We represent this transformation as a $2 \times 2$ matrix $M_i$ and we want to approximate this metric in the output $M^{out}$. The computation of $M_i$ is detailed in Section 5.1.

(2.2) In this phase, we compute the position of each vertex from the following equation.

$$E^k = \sum_{i=0} \sum_{j=1} w_{ij} \|e_{ij}^k - M_i^k e_{ij}^{std}\|^2, \quad (1)$$

where $e_{ij}^t, e_{ij}^{std}$ are edge vectors on the triangle $t_i^k$ and standard triangle $t_i^{std}$. We rewrite the function in terms of every edge vector:

$$E = \sum_t \sum_{i,j} w_{ij} \|(v_i^k - v_j^k) - M_{t_m}(v_i^{std} - v_j^{std})\|^2, \quad (2)$$

where each pair of $(v_i, v_j)$ belongs to the triangle $t_m$ (Note that $(v_i, v_j)$ and $(v_j, v_i)$ are 2 different vectors that belong to different triangles). $w_{ij}$ is the weight for each edge (see Paragraph "Weights" for details). Setting the gradient to zero, we obtain the following linear equation:

$$LV^{kT} = MLV^{std\,T}, \quad (3)$$

where the matrix L represents the edge relationship of vertices (weighted by $w_{ij}$). The matrix M includes all local matrix $M_{t_m}$, $V^k$ and $V^{std}$ are vectors including all vertices' positions on $M^k$ and standard triangles. $V^k$ is the only unknown vector here and solving this equation gives rise to the positions of all vertices in $V^k$.

Pre-factorization. We observe that the above matrix L depends only on the geometry of $M$. Thus this sparse matrix is fixed during iterations, allowing us to pre-factorize it with Cholesky decomposition and we can reuse the factorization many times throughout the algorithm in order to accelerate the process, which has a significant impact on algorithm efficiency. The total distortion error $E^k$ converges and we end the iteration when $\|E^k - E^{k-1}\|$ is smaller than the threshold α (we set α = 0.1%).

Weights. The choice of weight $w_{ij}$ in Eq.(2) depends on the importance of the triangle. The triangles around the ROI center are more sensitive to distortion. Meanwhile, the distortion on a large triangle is more visually confusing than that on the tiny ones. Therefore, we design the weight as $w_{ij} = (1 + h_m)A_m \cot(\theta)$, where $A_m$ is the area of the triangle $t_m$, $h_m$ is the averaged height (z-values) of the triangle, and θ is the opposite angle of the edge vector $(v_i, v_j)$ in $t_m$.

### 5.1 Computing Metrics

The vertex position in $M$ is determined by our designed metric $MI$. In our system, we want to achieve a flexible metric such that the user can generate variable visual effects with easy interaction. We notice that each transformation matrix includes two factors: one rotation matrix and two scaling values along two orthogonal directions. Inspired by [Liu et al. 2008], which blended the angle-only metric and rigid-only metric, we provide a new method that allows the user to specify a "mixed" metric that actually blends between two factors.

We start first by computing the transformation matrix between a triangle $t_i^k$ in $M^k$ and the standard triangle $t_i^{std}$. Equivalent to [Sorkine and Alexa 2007] and [Horn et al. 1988], we compute the Jacobian matrix **J** between two triangles.

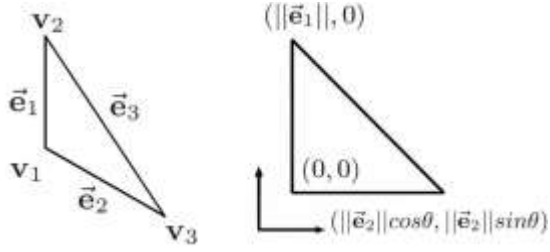

Figure 4: Generating a 2D standard triangle. Left: Original 3D triangle. Right: 2D standard triangle $t_i^{std}$.

$$J(t_i^k) = \sum_{i=1}^{X} e_i^k (e_i^{std})^T \qquad (4)$$

This matrix measures two tetrahedra's deformation on two factors: rotation and scaling. We can decompose two factors by singular value decomposition.

$$J = U\Sigma V^T, M_r = UV^T. \qquad (5)$$

Here $M_r$ is a rotation-only matrix. and $\Sigma$ includes two scaling values $\sigma_1$ and $\sigma_2$.

$$\Sigma = \begin{matrix} \sigma_1 & 0 \\ 0 & \sigma_2 \end{matrix}$$

To compute a flexible matrix, we can change this $2 \times 2$ diagonal matrix $\Sigma$ with blended scaling values. We allow the user to input a blending parameter $\alpha(0 \leq \alpha \leq 0.5)$. Then the resulting matrix is formulated as:

$$M = U \begin{matrix} \sigma_1^b & 0 \\ 0 & \sigma_2^b \end{matrix} V, \qquad (6)$$

where $\sigma_1^b = \alpha(\sigma_1 - 1) + 1, \sigma_2^b = \alpha(\sigma_2 - 1) + 1$.

## 6 Experimental Results

Our system can effectively provide F+C information to the user, allowing the user to get detailed focal region while maintaining the integral perception of the model. The results shown in the following figures demonstrate the power of our technique. Our experimental results are implemented on a 3GHz Pentium-IV PC with 4Giga RAM. In Fig. 5, we test our lens using several popular data structures such as graph, city, map, and text for information visualization: Graph is an abstract data structure representing relationships or connections. For access to relative nodes or to the particularly important nodes, our lens makes it easy to find and navigate toward these nodes; Our framework also improves the magnification functions with results of multi-scale map/satellite magnification, which reveal and magnify the additional details (e.g., additional country names); Our lens provides the efficient scanning function for the text reading as well. We can place the magnifier to zoom in the focus region while the remaining regions are evenly distributed to the context area (as shown in Fig. 1).

Algorithm 1 The flattening algorithm.

```
Input: triangle mesh M^3D,
       Blending parameter α ∈ [0, 0.5]
       Fitting error threshold
Output: 2D mesh M^out
L = BuildMatrix(M) // See Eq.(3)
Cholesky − Decomposition(L)
for all t_i^3D ∈ M^3D do
   //Compute the 2D standard triangle
   t_i^std = 2D − Standard(t_i^3D)
end for
Initial guess
M^0 = Projection(M^3D)
while ||E^k − E^(k−1)|| >   do
   for all t_i^k ∈ M^k do
      //Compute metrics. See Eq. (4)
      M_{t_i} = Compute(t_i^3D, t_i, α)
   end for
   // Build and solve Eq.(3) to get M^k
   E^t = FittingError(M^k, M^3D)
   k = k + 1
end while
M^out = M^k
```

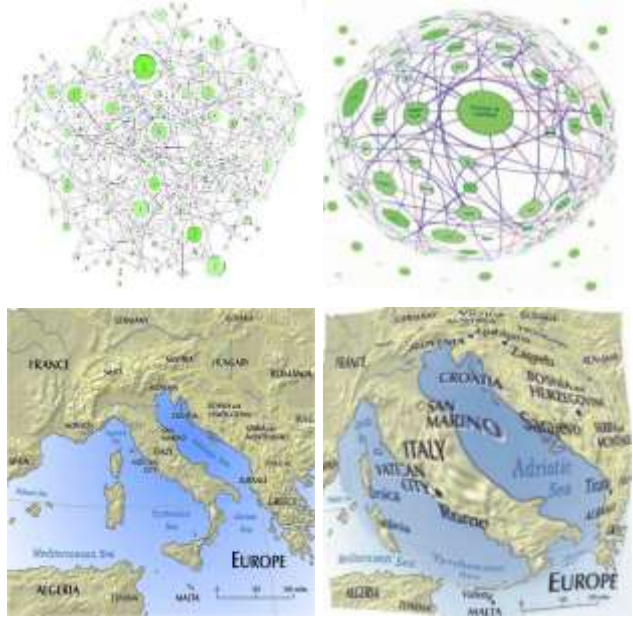

Figure 5: Applications of our lens simulation. Left row: Inputs. Right row: Graph of company relations, the connecting edges are revealed by the magnification; European map, major cities of Italy are revealed now.

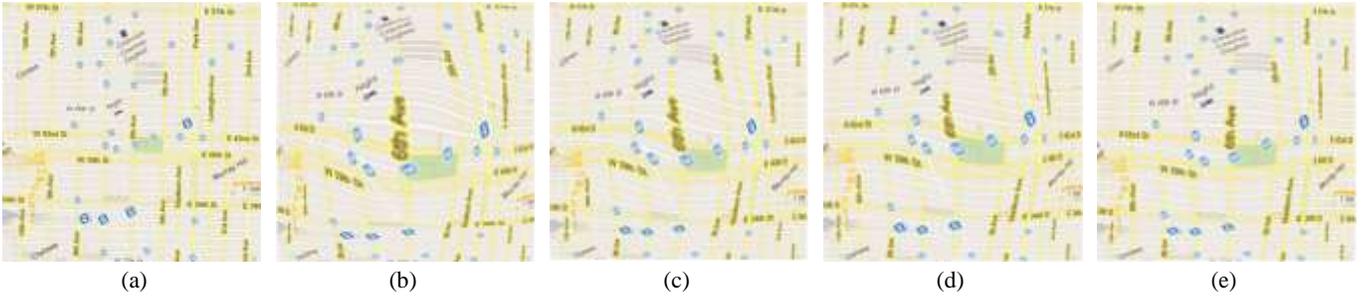

(a) (b) (c) (d) (e)

Figure 6: A group of different metrics with modified blending parameter α (α =0, 0.01 , 0.1, 0.5).

Fig. 2(d) is another excellent example to demonstrate that our technique offers a powerful lens for the route magnification. Using our lens, the user can see the additional route information and easily panning or zooming to achieve their requirements. Meanwhile, there is no any obviously visual distortion in both focus or context areas (the transition area with two view scales merges using linear interpolation). The global road distributions and orientations are preserved, and detailed streets are displayed around ROI.

As a general rule, a good F+C method should be able to maximally support the shape/feature preservation of objects of interest, such as conformal (angle) preservation or/and authalic (area) preservation, while minimizing context distortions [Zhao et al. 2011]. Instead of only minimizing angle distortion in [Zhao et al. 2012a, Zhao et al. 2012b,], Fig. 6 shows a group of lenses with the same input but different metrics, with the blending parameter α = 0, 0.01, 0.1, 0.5. This blend metrics en- rich the result and thus the user can modify the blending parameter to interactively change the visual effect until one result is satisfac- tory from the user's perspective.

Performance. Unlike other methods, the performance of our framework does not depend on the input image but the size of our triangle mesh. So a conventional performance table ("model-by-model") is not necessary for the analysis purpose. The sample images we tested are all between $512 \times 512$ and $1024 \times 1024$. We provide two meshes with sizes of $100 \times 100$ and $200 \times 200$ to handle small and large images separately. The smaller mesh (10k vertices) uses only 0.3 second for one iteration and it always converges in 2 iterations. We use the larger mesh (40k vertices) to handle very high-detailed application and it uses 1.3 seconds for one iteration. The pre-processing (matrix assembling and pre-factorization) requires only about 1.0 second.

Distortion. Similar to Eq.(2), we apply the following term to measure the shape distortion on every triangle $T_i$.

$$E_i = \sum_{j=1}^{X} w_{ij} \|e_j - M_i e_j^d\|^2, \quad (7)$$

Fig. 7 compares the distortion between our lens and poly-focal lens [Carpendale et al. 1997] (We consider the input image of poly-focal lens as a regular grid mesh. The deformation equation is defined in [Carpendale et al. 1997]). Although poly-focal lens or fisheye lens can have similar continuous magnification F+C view as our lens, it creates noticeable distortions towards its edges and has no method to formally control the focus region as well as to preserve local features in the context region. The comparison is meaningful because both methods allow "free-boundary" to obtain better shape-preserving effects. To measure the distortion of poly-focal lens, we also consider their resulting image as a deformed mesh with each vertex/color moving to the new position. Thus we can also use the same criteria to measure the shape distortion. The color indicates that our method can reduce the shape distortion in a much better way. We use blue color to represent zero distortion and red the maximum (0.45 in our result).

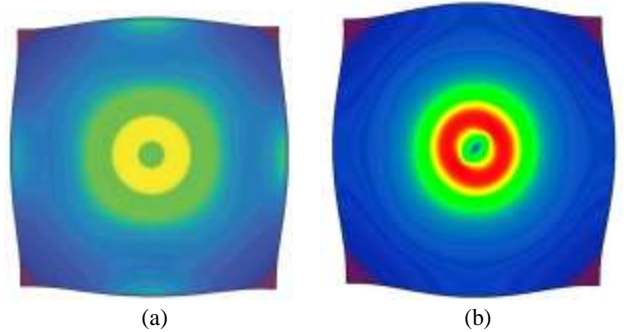

(a) (b)

Figure 7: (a-b) The distortion of our mesh and poly-focal lens. The distortion is color-coded from blue (minimum) to red (maximum).

Comparison for Magnification Results. We apply our method to a volumetric colon dataset to verify the advantages of our lens and compare with others as shown in Fig. 8. Local shape preservation and smooth transition have important applications in the clinic education, diagnose, and even virtual surgery. In the normal clinic exam, the colonoscopy needle navigates along the colon axis and the lens is added along the same direction such that the clinicians are able to recognize polyps on the folds (the wrinkles on the colon wall, red circle). The folds in Fig. 8(b-c) are seriously distorted which may sabotage the clinicians' expertise on polyps detection. No matter how we modify their lenses in (b-c), the distorted folds always exist along the lens boundary. In sharp contrast, the fold details in (d) are better preserved and easy for recognition.

We compare our method with other approaches, like zoom-in, fisheye, bi-focal, perspective wall, poly-focal [Carpendale et al. 1997] and cube deformation [Wang et al. 2008] in Table 1. Our method has advantages in the following aspects. First, our solution works well particularly with the complex shape, because it can flexibly design arbitrary shapes for lenses. Our method emphasizes angle and rigidity metrics for the shape-preserving purpose. Moreover, it allows the user to interactively design and blend various metrics.

Limitations. Our system flattens the mesh to achieve F+C visualization, but potentially it may result in flip-over phenomenon (i.e., the resulting triangle covers another one or its orientation is reversed). Fortunately, this phenomenon always happens especially on a highly curved surface with complex topology. In contrast, our 3D mesh is relatively very simple compared with common models used in geometric modeling study and there are no flip-over triangles in all examples during our experiments. The texturing step (Section 4.1) also produces a fine mapping as a good initial guess. Meanwhile, we can always solve the flip-over problem using the existing algorithm [Li et al. 2008].

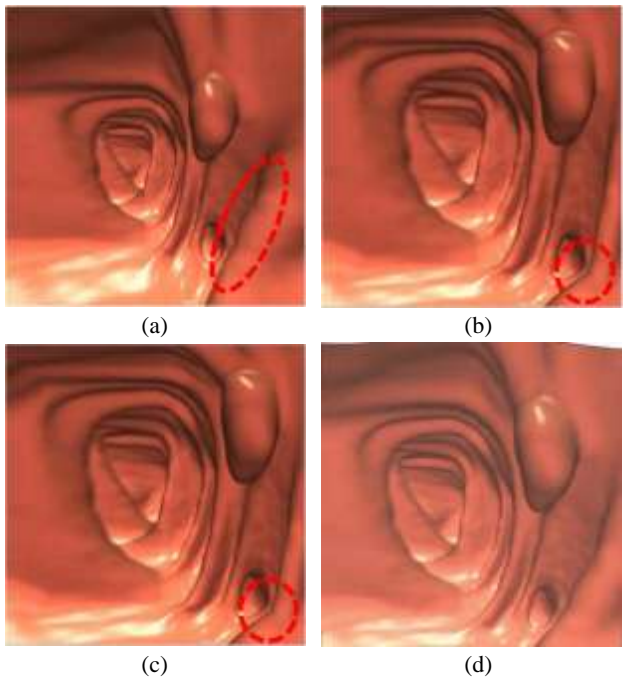

(a)            (b)

(c)            (d)

Figure 8: Magnification results using different lenses for volumetric colon dataset. (a) Original colon dataset. (b-d) Magnification results using bifocal, polyfocal, and our lenses. By comparison, the folds on the interior colon surface are seriously distorted by all the other lenses because of the sharp transition between the focus and context regions, while our lens shows the accurate shapes/features of the interior colon surface without any obvious distortion.

Compared with the direct zoom-in and bi-focal methods, our method can not authentically keep exactly the same feature of a local region as the original input. Also, our metric lacks of the measurement to preserve the global structure, shape symmetry, or long straight lines. However, our human cognitive system for recognition is accustomed to automatically compensating these slight variations of a local region and thus it relieves possible disdisturbing experience for the user.

## 7 Conclusion

We have developed a novel and interactive technique to achieve Focus+Context visualization based on geometric deformations. Specifically, we develop from the input a 3D mesh and magnify the ROIs through deformation on this mesh. Our lens design methodology and the prototype system manifest that the geometric deformation metrics greatly enhance the F+C visualization, and our approach is expected to transcend the traditional boundary of geometric modeling and will benefit data visualization and visual analytics [Li et al. 2012a, Li et al. 2012b].

The important features of our framework can be summarized as: (1) Shape-preserving. The geometric deformation metrics are minimized so that the resulting details appear similar to their original counterparts. Geometric deformation also generates a continuous transition region where the user can get a smooth viewing transition from the highly-magnified interior region to the non-magnified exterior region; (2) Robustness. It enables the user to design arbitrary number/shape of magnifiers to effectively display the entire ROIs for visualization of multiple and complex features. It also allows the user to interactively specify geometric metrics for various visual effects; (3) Efficiency. The computation is very efficient because of our pre-factorization processing. Our experimental results have demonstrated that our lens, as a novel F+C technique, has great potentials in many visualization applications.

In future work, we wish to explore the utility of 3D mesh in higher-dimensional datasets, inspired by [Li et al. preprint a, Li et al. preprint b]. We observe that the current scheme of 2D to 3D mesh deformation is capable of generalizing F+C visualization to handle more complicated datasets. The minimization of a linear system can be significantly accelerated on GPU platform. We also plan to enhance our work to handle more shape-sensitive applications (e.g., vascular system).

## References

BIER, E. A., STONE, M. C., PIER, K., BUXTON, W., AND DEROSE, T. D. 1993. Toolglass and magic lenses: The see-through interface. Computer Graphics, 73–80.

CARPENDALE, M. S. T., COWPERTHWAITE, D. J., AND FRACCHIA, F. D. 1997. Extending distortion viewing from 2D to 3D. IEEE Comput. Graph. Appl. 17, 4, 42–51.

CORREA, C. D., AND SILVER, D. 2007. Programmable shaders for deformation rendering. ACM SIGGRAPH/EUROGRAPHICS Symp. on Graphics Hardware, 89–96.

FLOATER, M. 2003. Mean value coordinates. Computer Aided Geometric Design 20, 1, 19–27.

FURNAS., G. 1986. Generalized fisheye views. Human Factors in Computing Systems, CHI Conference Proceedings, 16–23.

GANSNER, E., KOREN, Y., AND NORTH, S. 2004. Topological fisheye views for visualizing large graphs. IEEE Symposium on Information Visualization, 175–182.

HORN, B. K. P., HILDEN, H., AND NEGAHDARIPOUR, S. 1988. Closed-form solution of absolute orientation using orthonormal matrices. Journal of the Optical Society America 5, 7, 1127–1135.

JACOBSON, A., BARAN, I., POPOVIĆ, J., AND SORKINE, O. 2011. Bounded biharmonic weights for real-time deformation. ACM Transactions on Graphics (proceedings of ACM SIGGRAPH) 30, 4, 78:1–78:8.

KARNICK, P., CLINE, D., JESCHKE, S., RAZDAN, A., AND WONKA, P. 2010. Route visualization using detail lenses. IEEE Transactions on Visualization and Computer Graphics 16, 2, 235–247.

LAMAR, E., HAMANN, B., AND JOY, K. I. 2001. A magnification lens for interactive volume visualization. Pacific Conference on Computer Graphics and Applications, 223–232.

LI, X., BAO, Y., GUO, X., JIN, M., GU, X., AND QIN, H. 2008. Globally optimal surface mapping for surfaces with arbitrary topology. IEEE Transactions on Visualization and Computer Graphics 14, 4, 805–819.

Li, B. Li, X. Wang K. and Qin, H. Surface Mesh to Volumetric Spline Conversion with Generalized Polycube. IEEE Transactions on Visualization and Computer Graphics, in press.

Li, B. and Qin, H. Component-aware Tensor-product Trivariate Splines of Arbitrary Topology. IEEE Proceedings of Shape Modeling and Applications, 18:1–18:12, 2012.

Li, B. and Qin, H. Generalized PolyCube Trivariate Splines. IEEE Proceedings of Shape Modeling and Applications, 261-267, 2012

Li, B. Zhao, X and Qin, H . Four Dimensional Geometry Lens: A Novel Volumetric Magnification Approach. Computer Graphics Forum, pre-print.

LIPMAN, Y., LEVIN, D., AND COHEN-OR, D. 2008. Green coordinates. ACM Trans. Graph. 27 (August), 78:1–78:10.

LIU, L., ZHANG, L., XU, Y., GOTSMAN, C., AND GORTLER, S. J. 2008. A local/global approach to mesh parameterization. Computer Graphics Forum 27, 5, 1495–1504.

NEKRASOVSKI, D., BODNAR, A., MCGRENERE, J., GUIMBRETIÈRE, F., AND MUNZNER, T. 2006. An evaluation of pan & zoom and rubber sheet navigation with and without an overview. CHI Conference on Human Factors in Computing Sys- tems, 11–20.

PIETRIGA, E., AND APPERT, C. 2008. Sigma lenses: focus-context transitions combining space, time and translucence. Conference on Human Factors in Computing Systems, 1343–1352.

PIETRIGA, E., BAU, O., AND APPERT, C. 2010. Representation-independent in-place magnification with sigma lenses. IEEE Transactions on Visualization and Computer Graphics 16, 455–467.

PLAISANT, C., GROSJEAN, J., AND BEDERSON, B. B. 2002. Spacetree: Supporting exploration in large node link tree, design evolution and empirical evaluation. IEEE Symposium on Information Visualization, 57–64.

SCHAEFER, S., MCPHAIL, T., AND WARREN, J. 2006. Image deformation using moving least squares. ACM Trans. Graph. 25 (July), 533–540.

SHI, J., AND MALIK, J. 1997. Normalized cuts and image segmentation. In Proceedings of the 1997 Conference on Computer Vision and Pattern Recognition (CVPR '97), IEEE Computer Society, Washington, DC, USA, CVPR '97, 731– 740.

SHI, K., IRANI, P., AND LI, B. 2005. An evaluation of content browsing techniques for hierarchical space-filling visualizations. IEEE Symposium on Information Visualization, 11–18.

SORKINE, O., AND ALEXA, M. 2007. As-rigid-as-possible surface modeling. In Proceedings of the fifth Eurographics symposium on Geometry processing, Eurographics Association, Aire- la-Ville, Switzerland, Switzerland, 109–116.

STASKO, J., AND ZHANG, E. 2000. Focus+context display and navigation techniques for enhancing radial, space-filling hierarchy visualizations. IEEE Symposium on Information Visualization, 57–64.

STEGMAIER, S., STRENGERT, M., KLEIN, T., AND ERTL, T. 2006. A simple and flexible volume rendering framework for graphics-hardware-based raycasting. Fourth International Workshop on Volume Graphics, 187–241.

K. Wang, X. Li, B. Li, and H. Qin. Restricted Trivariate Polycube Splines for Volumetric Data Modeling. IEEE Transactions on Visualization and Computer Graphics, 18(5): 703-716, 2012.

WANG, Y.-S., LEE, T.-Y., AND TAI, C.-L. 2008. Focus+context visualization with distortion minimization. IEEE Transactions on Visualization and Computer Graphics 14, 1731–1738.

WANG, Y.-S., WANG, C., LEE, T.-Y., AND MA, K.-L. 2011. Feature-preserving volume data reduction and focus+context visualization. IEEE Transactions on Visualization and Computer Graphics 17, 2 (feb.), 171 –181.

YANG, J., WARD, M. O., RUNDENSTEINER, E. A., AND PATRO, A. 2003. Interring: a visual interface for navigating and manipulating hierarchies. IEEE Symposium on Information Visualization 2, 1, 16–30.

ZHAO, X., LI, B., WANG, L., AND KAUFMAN, A. 2011. Focus+context volumetric visualization using 3d texture-guided moving least squares. Proceedings of the Computer Graphics International.

ZHAO, X., LI, B., WANG, L., AND KAUFMAN, A. Texture-guided Volumetric Deformation and Visualization Using 3D Moving Least Squares. The Visual Computer, 28(2): 193-204, 2012.

ZHAO, X., ZENG, W., GU, X., KAUFMAN, A., XU, W., AND MUELLER, K. 2012. Conformal magnifier: A focus+ context technique with minimal distortion. IEEE Transactions on Visualization and Computer Graphics.

ZIEGLER, H., AND KEIM, D. A. 2008. Copernicus: Context-preserving engine for route navigation with interactive user-modifiable scaling. Comput. Graph. Forum 27, 3, 927–934.